\documentstyle[12pt]{article}
\newcommand{\be}{\begin{equation}}
\newcommand{\ee}{\end{equation}}
\newcommand{\ba}{\begin{eqnarray}}
\newcommand{\ea}{\end{eqnarray}}
\newcommand{\baa}{\begin{eqnarray*}}
\newcommand{\eaa}{\end{eqnarray*}}
\newcommand{\lab}[1]{\label{#1}}

\newcommand{\dis}{\displaystyle}

\newcommand{\bhat}{\hat{\beta}}
\begin{document}
{\pagestyle{empty}
\vskip 2.5cm
~\\

{\renewcommand{\thefootnote}{\fnsymbol{footnote}}
\centerline{\large \bf Molecular dynamics, Langevin, and hybrid
Monte Carlo}

\vskip 0.5cm
\centerline{\large \bf simulations in multicanonical ensemble} 
}
\vskip 3.0cm
 
\centerline{Ulrich H.E.~Hansmann,$^{a,}$
\footnote{\ \ e-mail: hansmann@ims.ac.jp} 
Yuko Okamoto,$^{a,}$ \footnote{\ \ e-mail: okamotoy@ims.ac.jp}
and Frank Eisenmenger$^{b,}$
\footnote{\ \ e-mail: eisenmen@orion.rz.mdc-berlin.de}} 
\vskip 1.5cm
\centerline{$^{a}$ {\it Department of 
Theoretical Studies, Institute for Molecular Science}}
\centerline{{\it Okazaki, Aichi 444, Japan}}
\vskip 0.5cm
\centerline{$^b${\it Institute for Biochemistry, Medical Faculty of
the Humboldt University Berlin}}
\centerline{{\it 10115 Berlin, Germany}} 

\medbreak
\vskip 3.5cm
 
\centerline{\bf ABSTRACT}
\vskip 0.3cm

We demonstrate that the multicanonical approach is not restricted
to Monte Carlo simulations, but can also be applied to 
 simulation techniques such as molecular dynamics, Langevin, and
hybrid Monte Carlo algorithms.  The effectiveness of 
the methods are tested with an energy function for the protein
folding problem.  Simulations in the multicanonical ensemble
by the three methods are performed for a penta peptide,
Met-enkephalin.  For each algorithm, it is shown that 
from only one simulation run one can
not only find the global-minimum-energy conformation 
but also obtain probability distributions in 
canonical ensemble at any temperature, which allows the calculation
of any thermodynamic quantity as a function of temperature.

\vfill
\newpage}
\baselineskip=0.8cm
\noindent
{\bf 1.~ INTRODUCTION} \\
Simulations in a system with many degrees of freedom by
conventional methods such as molecular dynamics (MD) and 
Monte Carlo (MC)
can sample only a small portion of the entire phase space,
rendering the calculations of various thermodynamic
quantities inaccurate.  
This is because the energy function has many local minima,
and 
at low temperatures  simulations will necessarily get trapped in the
configurations corresponding to one of these local minima.
In order to overcome this multiple-minima
problem, many methods have been proposed.  For instance,
simulated annealing \cite{SA} is one of the most widely
used algorithms to locate the global-minimum state
out of the multitude of local-minimum states.
The multicanonical approach \cite{MU,MU3} is another 
powerful technique.  The advantage of this algorithm lies in
the fact that from only one simulation run one can not only 
find the energy global minimum but also calculate
various thermodynamic quantities
at any temperature.  The method was originally developed
to overcome the supercritical slowing down of first-order
phase transitions,\cite{MU,MU3} and it was then 
proposed to be used for systems that suffer from the multiple-minima
problem such as spin glasses \cite{BC} and the protein
folding problem.\cite{HO}
The same method was later referred to as entropic sampling,\cite{ES}
but the proof of the equivalence of the two methods was given
to clarify the matter.\cite{BHO} 
In the context of the protein
folding problem, the effectiveness of multicanonical algorithms 
was compared with that of simulated annealing.\cite{HO2} 
It was also used to study the 
coil-globular transitions of a model protein,\cite{HS}
helix-coil transitions of amino-acid homo-oligomers,\cite{HO3}
and conformational sampling of a constrained peptide.\cite{K}

In all of the previous works the multicanonical ansatz was used
in the context of Monte Carlo simulations utilizing mostly the 
Metropolis algorithm \cite{Metro} to generate a Markov chain 
of configurations. 
However, 
other simulation techniques such as
 molecular dynamics~\cite{MD} are 
also widely used.
The purpose of the present work is to 
demonstrate that these techniques 
can be used for simulations in {\it multicanonical ensemble}.
Here, we consider three common algorithms:
molecular dynamics, Langevin,\cite{Lang} and hybrid Monte 
Carlo.\cite{HMC}
The performance of the algorithms are tested with the system
of an oligopeptide, Met-enkephalin.\\

\noindent
{\bf 2.~ METHODS}\\
{\bf {\it 2.1.~ Multicanonical ensemble}}\\
Simulations in the canonical ensemble at temperature $T$ 
weigh each state with the
Boltzmann factor
\begin{equation}
w_{B}(E,T) =  e^{-\hat{\beta} E}~,
\label{eq0}
\end{equation}
where the inverse temperature is given by $\hat{\beta} = \frac{1}{k_BT}$  
with Boltzmann
constant $k_B$. This weight factor gives the usual bell-shaped
canonical probability distribution of energy:
\begin{equation}
P_B(E,T) \propto n(E)~w_B (E,T)~,
\label{eq1}
\end{equation}
where $n(E)$ is the density of states.
 
In the {\it multicanonical ensemble},\cite{MU} on the other hand, 
the probability
distribution of energy is {\it defined} to be constant:
\begin{equation}
 P_{mu}(E) \propto n(E)~w_{mu}(E) = {\rm const.}
\label{eq2}
\end{equation}
The multicanonical weight factor for each state with
energy $E$ is then given by
\begin{equation}
 w_{mu} (E)\propto n^{-1}(E) = e^{-S(E)}~,
\label{eq3}
\end{equation}
where $S(E)$ is the microcanonical entropy (with $k_B = 1$):
\begin{equation}
S(E) = \ln n(E)~.
\label{eq4}
\end{equation}
With the uniform probability distribution of Eq.~(\ref{eq2}), a simulation
in multicanonical ensemble leads to a 1D random walk in energy space,
allowing itself to escape from any energy barrier and to explore
wide range of the phase space.

Unlike in a canonical simulation, however, the multicanonical weight 
$w_{mu} (E)$ is not {\it a priori} known,
and one has to obtain its estimator for a numerical 
simulation.  Hence, the multicanonical ansatz consists of
three steps:
In the first step the estimator of the multicanonical weight factor
$w_{mu} (E)$ is calculated (for details of the method of finding
$w_{mu} (E)$ for the case of Metropolis Monte Carlo algorithm, see 
Refs.~\cite{MU3,HO2}).
Then one makes  with this weight factor a production run
with high statistics. In this way information  is collected over 
the whole energy range. 
Finally, by examining the history of this simulation, one can not
 only locate the energy global minimum but also obtain the
 canonical distribution at any inverse temperature $\hat{\beta}$
 for a wide
 range of temperatures by the re-weighting techniques:\cite{FS}
\begin{equation}
 P_B(E,T) \propto P_{mu} (E)~w^{-1}_{mu} (E)~e^{-\hat{\beta} E}~.
\label{eq5}
\end{equation}
This allows one to calculate the expectation value of any physical
quantity ${\cal O}$ by
\begin{equation}
< {\cal O} >_T ~= \frac{\displaystyle{\int dE~ {\cal O} (E) P_B(E,T)}}
                       {\displaystyle{\int dE~ P_B(E,T)}}~.
\label{eq6}
\end{equation}

In the following subsections, we describe how to implement
multicanonical simulations for Langevin, molecular dynamics, 
and hybrid Monte Carlo algorithms.\\
\noindent
{\bf {\it 2.2.~ Langevin algorithm in multicanonical ensemble}}\\
The Langevin algorithm\cite{Lang} is used to integrate
 the following  differential equation:
\be
\dot{q}_i = -  \bhat \dis{\frac{\partial E(q)}{\partial q_i}} + \eta_i~, 
\lab{eq10}
\ee
where  $q_i \ (i=1,\cdots,N)$ are the (generalized) coodinates of the system,
$E(q)$ is the potential energy,
and $\eta_i$ is a set of independent Gaussian distributed
random variables with a unit variance:  
\be
< \eta_i (t_l) \eta_j (t_m) > = \delta_{ij} \delta(t_l - t_m).
\lab{eq10p}
\ee
It can be shown that 
the dynamics based on the Langevin algorithm yields a canonical
distribution $P_B (E,T) \propto n(E) e^{-\hat{\beta}E}$.  For
numerical work one
integrates the above equation by discretizing 
the time with step $\Delta t$~:
\be
q_i(t+\Delta t) = q_i(t) + \Delta t
\left( - \bhat \dis{\frac{\partial E(q)}{\partial q_i(t)}} + \eta_i(t)
\right)~. \\
\lab{eq11}
\ee

A straightforward generalization of this technique to simulations in 
multicanonical ensemble can be  made by
replacing the $\bhat E$ in Eq.~(\ref{eq10})
by the microcanonical entropy $S(E)$: 
\be
\dot{q}_i = - \dis{\frac{\partial S(E(q))}{\partial q_i}} + \eta_i~. 
\lab{eq10a}
\ee
The above equation now describes a dynamics which will yield a 
{\it multicanonical} distribution $P_{mu} (E) \propto n(E) e^{-S(E)} 
= {\rm const.}$ (see Eq.~(\ref{eq3})).  
(A similar consideration
of multicanonical Langevin algorithm is given in Ref.~\cite{MO}.)
Hence, for actual 
simulations we use the following difference equation:
\be
q_i(t+\Delta t) = q_i(t) + \Delta t
\left( - \dis{\frac{\partial S(E(q))}{\partial q_i(t)}} + \eta_i(t)
\right)~. \\
\lab{eq11a}
\ee

We remark that Eq.~(\ref{eq10a}) can be written as
\be
\dot{q}_i = - \dis{\frac{\partial S}{\partial E}}
\dis{\frac{\partial E(q)}{\partial q_i}} + \eta_i 
 = - \beta (E) \dis{\frac{\partial E(q)}{\partial q_i}} + \eta_i~, 
\lab{eq11b}
\ee
where $\beta (E)$ is an energy-dependent effective inverse temperature.
In this notation the term ``multicanonical'' becomes obvious
(compare Eq.~(\ref{eq11b}) with Eq.~(\ref{eq10})).

\noindent
{\bf {\it 2.3.~ Molecular dynamics algorithm in multicanonical 
ensemble}}\\
The expectation value of a physical quantity ${\cal O}$ is
calculated by
\be
< {\cal O} >_T ~
= \frac{\dis{\int Dq~ {\cal O} (q) e^{-\bhat E(q)}}}
{\dis{\int Dq~ e^{-\bhat E(q)}}}~,
\label{eq7p}
\ee
where the integration measure is defined by $Dq  = \prod_{i=1}^N dq_i$
and $q_i \ (i=1,\cdots,N)$ are again the (generalized) coordinates of
a system. $E(q)$ is the potential energy of the system. The above
equation is mathematically identical with 
\be
< {\cal O} >_T ~
= \frac{\dis{\int} DqD\pi ~
{\cal O} (q) {\rm exp} \left( -\dis{\sum_{i=1}^{N}} \dis{\frac{\pi_i^2}{2m_i}}
 -\bhat E(q) \right)}
{\dis{\int} DqD\pi~ {\rm exp} \left( -\dis{\sum_{i=1}^{N}} 
\dis{\frac{\pi_i^2}{2m_i}}
 -\bhat E(q) \right)}~,
\label{eq7pa}
\ee
where we used the notation $D\pi = \prod_{i=1}^N d\pi_i$. Identifying
the auxillary variables $\pi_i$ with the conjugate momenta corresponding to
the coordinates $q_i$, we can describe our system with a Hamiltonian
\be
H(q,\pi) = \frac{1}{2} \sum_{i=1}^N \pi_i^2 + \hat{\beta}E(q_1,\cdots,q_N)~,
\lab{eq7}
\ee
where we have set all the masses $m_i$ equal to 1 for
simplicity.

The classical molecular dynamics algorithm uses    
the Hamilton's equations of motion
\be
\left\{
\begin{array}{rl}
\dot{q}_i &= \dis{\frac{\partial H}{\partial \pi_i}} = \pi_i~,\\ 
\dot{\pi}_i &= - \dis{\frac{\partial H}{\partial q_i}} 
           = - \bhat \dis{\frac{\partial E}{\partial q_i}}~,
\end{array}
\right.
\lab{eq8}
\ee
to generate representative ensembles of configurations.
For numerical work  the time is discretized with step $\Delta t$
and the equations are integrated according to the {\it leapfrog}
(or other time reversible integration) scheme:
\be
\left\{
\begin{array}{rl}
q_i(t+\Delta t) &= q_i(t) + \Delta t~ 
\pi_i \left( t + \dis{\frac{\Delta t}{2}} \right)~,\\ 
\pi_i \left( t+\dis{\frac{3}{2}\Delta t} \right) 
&= \pi_i \left( t+\dis{\frac{\Delta t}{2}} \right) - \Delta t~
\bhat \dis{\frac{\partial E}{\partial q_i(t+\Delta t)}}~.
\end{array}
\right.
\lab{eq12}
\ee
The initial momenta 
$\{\pi_i(\frac{\Delta t}{2})\}$ for the iteration are
prepared by
\be
\pi_i \left( \dis{\frac{\Delta t}{2}} \right) 
= \pi_i(0) - \dis{\frac{\Delta t}{2}}
\bhat \dis{\frac{\partial E}{\partial q_i(0)}}~,
\lab{eq13}
\ee
with appropriately chosen  $q_i(0)$ and $\pi_i (0)$
($\pi_i (0)$ is from a Gaussian distribution).

In order to generalize this widely used technique to simulations 
in multicanonical
ensemble, we  again propose to replace $\bhat E$ by the
entropy $S(E)$ in Eqs.~(\ref{eq8}), (\ref{eq12}), and (\ref{eq13})
 (just as we
did for the Langevin algorithm). Hence, we have a new
\lq \lq Hamiltonian"
\be
H(q,\pi) = \frac{1}{2} \sum_{i=1}^N \pi_i^2 + S(E(q))~,
\lab{eq7a}
\ee
and a new set of Hamilton's equations of motion
\be
\left\{
\begin{array}{rl}
\dot{q}_i &= \dis{\frac{\partial H}{\partial \pi_i}} = \pi_i~, \\ 
\dot{\pi}_i &= - \dis{\frac{\partial H}{\partial q_i}} 
            = - \dis{\frac{\partial S(E(q))}{\partial q_i}} 
 = - \dis{\frac{\partial S}{\partial E}}
\dis{\frac{\partial E(q)}{\partial q_i}}~. 
\end{array}
\right.
\lab{eq8a}
\ee
This is the set of equations we adopt for multicanonical MD
simulations.  Formally it can be understood as a rescaling of the usual
 force term  by the derivative of the entropy. For numerical simulations
the Hamilton equations are again discretized in time and integrated by
a {\it leapfrog} scheme.\\
\noindent
{\bf {\it 2.4.~ Hybrid Monte Carlo algorithm in multicanonical 
ensemble}}\\
The hybrid Monte Carlo algorithm\cite{HMC} is based on the 
combination of molecular dynamics and Metropolis Monte Carlo
algorithms.  Namely, each proposal  for the 
Metropolis method is prepared by a short MD run starting from the actual
configuration.  Hence, this
algorithm is based on a global update, while in the 
conventional Metropolis 
method one is usually restricted to a local update. Furthermore, 
the Metropolis
step ensures that the sampled configurations are distributed according to
the chosen ensemble, while  convential molecular dynamics simulations are
hampered by difficult-to-control systematic errors 
due to 
finite step size in the
integration of the equations of motion.

Given the set of coordinates $\{q_i\}$ of the previous configuration and
choosing the corresponding momenta $\{\pi_i\}$ from a Gaussian
distribution, a certain number of MD steps are performed  
to obtain a candidate configuration $\{q_i^{\prime},\pi_i^{\prime}\}$.
This candidate is accepted according to the
Metropolis Monte Carlo criterion with probability
\be
p = \min \{ 1, e^{- (H(q^{\prime},\pi^{\prime}) - H(q,\pi))} \}~,
\lab{eq14}
\ee
where $H$ is the Hamiltonian in Eq.~(\ref{eq7}). The time reversibility of the
{\it leapfrog} integration scheme ensures detailed balance and therefore convergence
to the correct distribution.
The whole process is repeated for a desired number of times (Monte 
Carlo steps). The number of integration ({\it leapfrog}) steps $N_{LF}$ and 
the size of the time step $\Delta t$ are free parameters in the hybrid 
Monte Carlo
algorithm, which have to be tuned carefully. A choice of 
too small $N_{LF}$ and $\Delta t$
means that the sampled configurations are too much correlated,
while too large $N_{LF}$ (or $\Delta t$) yields high rejection rates. In both cases
the algorithm becomes inefficient.

The  generalization of this technique to simulations in multicanonical
ensemble can again be made by replacing the Hamiltonian of Eq.~(\ref{eq7})
with the multicanonical Hamiltonian of Eq.~(\ref{eq7a}), i.e.,~replacing
 $\hat{\beta} E$  
 by the entropy $S(E)$ in the equations of
motion.\\

\noindent
{\bf 3.~ RESULTS AND DISCUSSION}\\
The effectiveness of the algorithms presented in the previous 
section is tested for the system of an oligopeptide, Met-enkephalin.
This peptide has the amino-acid sequence Tyr-Gly-Gly-Phe-Met.
The potential energy function that we used is given 
by the sum of 
electrostatic term, Lennard-Jones term, and
hydrogen-bond term for all pairs of atoms in the peptide
together with the torsion term for all torsion angles.
The parameters for the energy function were adopted from
ECEPP/2.\cite{EC1}-\cite{EC3}  The computer code SMC \cite{SMC} was
modified to accomodate the multicanonical ensemble.  

For the coordinates $\{q_i\}$ we used the dihedral angles. (
We remark that it was recently
claimed that  convergence is faster for the dihedral 
coordinates.\cite{HMC2}  
Of course we 
could have used Cartesian coordinates as well 
with the same set of equations.) 
The peptide-bond dihedral angles $\omega$ were fixed to be
180$^{\circ}$ for simplicity.  This leaves 19 
dihedral angles as generalized coordinates.
By definition of multicanonical ensemble, one cannot obtain information
on the real dynamics of the system by the MD algorithm, and
only static thermodynamic quantities can be calculated.
For this reason we do not need to consider the 
equations of motion for dihedral space
as presented in Ref.~\cite{MDA}, but can use the much simpler form as given in
the previous section. However, we remark
that this may not be the optimal choice. Very often it
may be more suitable to distinguish between ``soft'' and ``hard''
degrees of freedom and introduce appropriately chosen ``masses'' in
the equations of motion.\cite{HMC2}

For the multicanonical MD simulations, we made a single
production run with the total number of time 
steps $N_{LF} = 400,000*19$ and the 
time-step size $\Delta t = 0.005$ (in arbitary units), after the
optimal estimate for the multicanonical weight factor $w_{mu}(E)$, 
or entropy $S(E)$, was obtained.  For the multicanonical
Langevin algorithm,
a production run with 
the same number of time steps 
($N_{LF} = 400,000*19$) as in
the MD simulation, but our optimal  
time-step size was only $\Delta t = 0.0001$.  This 
indicates that the simulation moves more slowly
through phase space, and we expect slower convergence
to the multicanonical distribution than in MD case.
For the multicanonical 
hybrid Monte Carlo algorithm, an MD simulation with 
19 leapfrog steps was made for each Monte Carlo step and
a production run with 200,000 MC steps was made. Since the Metropolis step 
in hybrid Monte Carlo corrects for errors due to the numerical integration
of the equation of motion, the time-step size can be large for 
this algorithm. We
chose $\Delta t = 0.01$ in our units. The initial 
conformation
for all three simulations was the final (and therefore equilibrized) 
conformation obtained from
a multicanonical Monte Carlo simulation of 200,000 sweeps, 
following 1,000 sweeps for thermalization with the same weights (in
each sweep all of the 19 angles were updated once). 

In Fig.~1 the time series of the total potential energy are
shown for the three multicanonical simulations.  They all
display a random walk in energy as they should for a simulation
in multicanonical ensemble.  All the lowest-energy conformations 
were essentially the same (with only a small
amount of deviations for each dihedral angle) as that of the 
global-minimum
energy conformation previously obtained for the same
energy function (with $\omega = 180^{\circ}$) by other 
methods.\cite{OKK,HO,MMV}  The global-minimum potential energy value 
obtained by
minimization is $-10.7$ kcal/mol.\cite{MMV} 
The random walks of the MD and hybrid MC simulations visited
the global-minimum region ($E < -10$ kcal/mol) three times 
and five times, respectively, while
that of the Langevin simulation reached the region only once.
These visits are separated by the walks towards the high energy region
much above $E = 16$ kcal/mol, which corresponds to the average energy 
at $T=1000$ K.\cite{HO} 

In Fig.~2a the time series of the end-to-end distance $r$ is plotted.
Here, the distance was measured from N of Tyr 1 to O of Met 5.
Only the result from the multicanonical hybrid Monte Carlo simulation
is given, since the other two simulations give similar results.
Note that there is a positive correlation between potential energy $E$
and end-to-end distance $r$ (compare Figs.~1c and 2a), indicating that 
a folded structure generally has a lower potential energy than a
stretched one. This becomes even clearer in Fig.~2b, where we display
the average end-to-end distance $r$ as a function of potential energy $E$.

In Fig.~3 we demonstrate that the probability 
distribution $P_{mu}(E)$ of potential
energy $E$ obtained from the multicanonical MD simulation    
is essentially flat (of the same order of magnitude) over the whole
energy range.  Similar figures can be drawn for the other two
algorithms.
 
In Fig.~4 the entropy $S(E)$ calculated  from the probability
distribution $P_{mu}(E)$ is displayed (see Eqs.~(\ref{eq2}) 
and (\ref{eq3})).  Only the result from
multicanonical MD simulation is given, since the other two simulations
give essentially the same results.  It is a monotonically increasing
function.  Note that there is a sudden drop of $S(E)$ near 
$E=-10$ kcal/mol, suggesting that the global-minimum conformation
is \lq \lq unique".

Simulations in multicanonical ensemble can not only find the energy
global minimum but also any thermodynamic quantity as a function
of temperature from a single simulation run.  We have calculated
the specific heat and average potential energy as functions of
temperature for the three algorithms.  The results all agreed
within errors  
with those from our previous multicanonical MC runs (see,
for instance, Refs.~\cite{HO,HO2}). 
Here, we just show another
example of such a calculation, the average end-to-end distance
as a function of temperature.  The results are essentially the
same for the three algorithms.  That from multicanonical Langevin algorithm 
is shown in Fig.~5.  We see that the average end-to-end distance
becomes smaller as the temperature is lowered, indicating that
the peptide has a compact structure at low temperatures.\\

\noindent
{\bf CONCLUSIONS} \\
In this article we have  shown that the  multicanonical ansatz is
not restricted to Monte Carlo simulations, but can also be used in 
combination with other simulation methods such as molecular 
dynamics, Langevin, and hybrid Monte Carlo algorithms.
We have tested the performances of these three methods in multicanonical
ensemble for a simple peptide, Met-enkephalin.  The results were
comparable to those of the original Monte Carlo version.\cite{HO}  
We believe that there is a wide range of applications for multicanonical
versions of molecular dynamics and related algorithms.  
For instance, multicanonical MD simulations may prove to be a valuable tool 
 for refinement of the protein structures
inferred from X-ray and/or NMR experiments.  

\vspace{0.5cm}
\noindent
{\bf Acknowledgements}: \\
The authors thank A. Kidera and N. Nakajima for informing us that they 
have also developed a method for
implementing an MD algorithm in multicanonical ensemble.
We are grateful to F. Hirata for letting us know the existence of
Refs.~\cite{MO,MDA}.
Our simulations were performed on the computers of the Computer
Center at the Institute for Molecular Science, Okazaki,
Japan.  This work is
supported, in part, by the Grants-in-Aid for Scientific Research
from the
Japanese Ministry of Education, Science, Sports, and Culture.\\


\newpage
\noindent
{\bf \Large FIGURE CAPTIONS:}\\
FIG.~1. (a) Time series of the total potential energy $E$ from a 
multicanonical Langevin simulation of
        400,000*19 time steps with step size $\Delta t = 0.0001$.
(b) Time series of $E$ from a multicanonical molecular dynamics  
simulation of
        400,000*19 time steps with step size $\Delta t = 0.005$.
(c) Time series of $E$ from a multicanonical hybrid Monte Carlo simulation of
        200,000 MC steps.  For each MC step an MD run of 19 
time steps was made with step size $\Delta t = 0.01$. \\
\noindent
FIG.~2. (a) Time series of end-to-end distance $r$
        from the multicanonical hybrid Monte Carlo simulation.
(b)  The average end-to-end distance $r$ as a function of 
potential energy $E$ obtained
        from the multicanonical hybrid Monte Carlo simulation. \\
\noindent
Fig.~3.  Probability distribution of potential energy $E$ obtained 
from the multicanonical molecular dynamics 
         simulation. \\
\noindent
Fig.~4:  Microcanonical entropy $S(E)$ as a function of potential energy $E$
         obtained from the multicanonical molecular dynamics simulation.\\
\noindent
FIG.~5:  The average end-to-end distance $r$ as a function of 
temperature obtained
         from the multicanonical Langevin simulation.

\noindent
\begin{thebibliography}{(00)}
\bibitem{SA} S. Kirkpatrick, C.D. Gelatt, Jr., and M.P. Vecchi,  
  {\it Science} {\bf 220} (1983) 671.
\bibitem{MU} B.A. Berg and T. Neuhaus, {\it Phys. Lett.} {\bf B267}
  (1991) 249; {\it Phys. Rev. Lett.} {\bf 68} (1992) 9.
\bibitem{MU3} B.A. Berg, {\it Int.~J.~Mod.~Phys.} {\bf C3}
  (1992) 1083.
\bibitem{BC} B.A. Berg and T. Celik, {\it Phys. Rev. Lett.} {\bf 69}
  (1992) 2292; B.A. Berg, U.H.E. Hansmann, and T. Celik,
{\it Phys. Rev.} {\bf B50} (1994) 16444.
\bibitem{HO} U.H.E. Hansmann and Y. Okamoto, {\it J.~Comp.~Chem.} 
   {\bf 14} (1993) 1333.
\bibitem{ES} J. Lee, {\it Phys.~Rev.~Lett.} {\bf 71} (1993) 211.
\bibitem{BHO} B.A. Berg, U.H.E. Hansmann, and Y. Okamoto, 
  {\it J. Phys. Chem.} {\bf 99} (1995) 2236.
\bibitem{HO2} U.H.E. Hansmann and Y. Okamoto, {\it J. Phys. Soc. Jpn.}
  {\bf 63} (1994) 3945; {\it Physica A} {\bf 212} (1994) 415.
\bibitem{HS} M.H. Hao and H.A. Scheraga, {\it J. Phys. Chem.} 
  {\bf 98} (1994) 4940.
\bibitem{HO3} Y.~Okamoto, U.H.E.~Hansmann, and T. Nakazawa,\ 
 {\it Chem. Lett.} {\bf 1995} 391;
 Y.~Okamoto and U.H.E.~Hansmann,\ {\it J.~Phys.~Chem.}
		{\bf 99} (1995) 2236.
\bibitem{K} A. Kidera, {\it Proc.~Nat.~Acad.~Sci.~U.S.A.} {\bf 92}
  (1995) 9886.
\bibitem{Metro} N. Metropolis, A.W. Rosenbluth, M.N. Rosenbluth, 
  A.H. Teller, and E. Teller, {\it J. Chem. Phys.} {\bf 21}
  (1953) 1087. 
\bibitem{MD} For instance, L. Verlet, {\it Phys. Rev.} {\bf 159} (1967) 98.
\bibitem{Lang} G. Parisi and Y.-S. Wu, {\it Sci. Sin.} {\bf 24} (1981) 483.
\bibitem{HMC} S. Duane, A.D. Kennedy, B.J. Pendleton, and D. Roweth,
 {\it Phys. Lett.} {\bf B195} (1987) 216.
\bibitem{FS} A.M. Ferrenberg and R.H. Swendsen, {\it Phys.\ Rev.\ Lett.}
  {\bf  61} (1988) 2635; {\it ibid.} {\bf 63 } (1989) 1658(E), and
  references given in the erratum.
\bibitem{MO} T. Munakata and S. Oyama, \lq \lq Adaptation and
linear response theory", Kyoto University preprint.
\bibitem{EC1} F.A. Momany, R.F. McGuire, A.W. Burgess, and H.A.
Scheraga, {\it J. Phys. Chem.} {\bf 79} (1975) 2361.
\bibitem{EC2} G. N{\'e}methy,
M.S. Pottle, and H.A. Scheraga, {\it J. Phys. Chem.} {\bf 87} (1983)
1883.
\bibitem{EC3} M.J. Sippl, G. N{\'e}methy, and H.A. Scheraga,
{\it J. Phys. Chem.} {\bf 88} (1984) 6231.
\bibitem{SMC} The program SMC was written by F. Eisenmenger.
\bibitem{HMC2} B.M. Forrest and U.W. Suter,
 {\it J. Chem. Phys.} {\bf 101} (1994) 2616.
\bibitem{MDA} A.K. Mazur, V.E. Dorofeev, and R.A. Abagyan,
{\it J. Comp. Phys.} {\bf 92} (1991) 261.
\bibitem{OKK} Y.~Okamoto, T.~Kikuchi, and H.~Kawai, {\it Chem. Lett.}
{\bf 1992} 1275.
\bibitem{MMV} H. Meirovitch, E. Meirovitch, A.G. Michel, 
and M. V{\'a}squez, {\it J. Phys. Chem.} {\bf 98} (1994) 6241.
   
\end{thebibliography}
\end{document}